\documentstyle[12pt,epsfig]{article}

\begin{document}
\begin{titlepage}
\begin{flushright}
UCLA/95/TEP/3\\
hep-ph/9502230 \\
January 29, 1995 \\
\end{flushright}
\vskip 2.cm
\begin{center}
{\Large\bf Second Order Fermions in Gauge Theories}
\vskip 2cm
{\large  A.G. Morgan\footnote{
Email address: {\tt morgan@physics.ucla.edu}
}}
\vskip 0.5cm
{\it Department of Physics, University of California at Los Angeles, \\
Los Angeles, CA 90095-1547 }
\vskip 3cm
\end{center}
\begin{abstract}
The second order formalism for fermions provides a description of
fermions that is very similar to that of scalars.  We demonstrate that
this second order formalism is equivalent to the standard Dirac
formalism. We do so in terms of the conventional fermionic Feynman
rules. The second order formalism has previously proven useful for the
computation of fermion loops, here we describe how the corresponding
rules can be applied to the calculation of amplitudes involving
external fermions, including tree-level processes and processes with
more than one fermion line. We comment on the supersymmetric
identities relating fermions and scalars and the associated
simplifications to perturbative calculations that are then more
transparent.
\end{abstract}
\vfill
\end{titlepage}

\def\eqn#1{Eq.~(\ref{#1})}
\def\eqns#1{Eqs.~(#1)}
\def\fig#1{Fig.~\ref{#1}}
\def\figs#1{Figs.~#1}
\def\Section#1{}
\def\Tr{{\rm Tr}}
\def\bra#1{\left\langle #1 \right|}
\def\ket#1{\left| #1 \right\rangle}
\def\braket#1#2{\left\langle #1 |\  #2 \right\rangle}

\newbox\charbox
\newbox\slabox
\def\s#1{{      
        \setbox\charbox=\hbox{$#1$}
        \setbox\slabox=\hbox{$/$}
        \dimen\charbox=\ht\slabox
        \advance\dimen\charbox by -\dp\slabox
        \advance\dimen\charbox by -\ht\charbox
        \advance\dimen\charbox by \dp\charbox
        \divide\dimen\charbox by 2
        \raise-\dimen\charbox\hbox to \wd\charbox{\hss/\hss}
        \llap{$#1$}
}}

\Section{Introduction}

The second order formalism for fermions provides a description of
fermions that is very similar to that of scalars. The prefix second
being due to the order of the derivative in the associated effective
action---the $\sim 1/k^2$ form of the propagator. Its has been
obtained from an analysis of the infinite string tension limit of
string theories \cite{BernKosower,BernDunbar,Lam} and is related to
the first quantized form \cite{Polyakov,Strassler}. It has
also been used in perturbative calculation \cite{BernEtAl}, where it
simplified the calculation of one loop amplitudes in QCD, gravity and
the electroweak sector of the Standard model.

It is the purpose of this letter to make clear the relationship
between the first and second order fermion rules, in a manner that
will be immediately obvious to Feynman diagram calculators. Further we
generalize the {\em fermion-loop} inspired rules to include a
description of external fermions.

By way of an example, we shed some light on the supersymmetric
similarity of scalars and fermions interacting with gauge fields. With
respect to the mechanics of calculating Feynman diagrams this has
previously been quite mysterious.

\Section{A formal derivation}

For completeness, we begin by reviewing the {\em formal} approach to
the derivation of the second order formalism. The basic idea being to
exploit the relationship between fermionic actions and straightforward
determinants. This is completely analogous to the derivation of the
Faddeev-Popov ghost effective Lagrangian term \cite{FaddeevPopov},
which is required for the calculation of general non-abelian
amplitudes.

We choose the following convention for our gauge coupling,
\begin{equation}
D_\mu = \partial_\mu + i g Q A_\mu .
\end{equation}
For QED the coupling is $g = e$, and the associated charge is $Q = -1$
for the electron. In QCD, and related non-abelian theories, the
coupling is required to be universal, $Q = 1$, and the gauge field is
expanded in terms of hermitian generators for the gauge symmetry,
$A_\mu = T^a A_\mu^a$. With these conventions we define the field
strength tensor to have the form,
\begin{equation}
F_{\mu \nu} = \frac{1}{i g Q} \left[ D_\mu, D_\nu \right]
	= \left( \partial_\mu A_\nu \right)
	- \left( \partial_\nu A_\mu \right)
	+ i g Q \left[ A_\mu , A_\nu \right].
\end{equation}

The starting point is the generating functional for the effective
action \cite{EffectiveAction} of a closed fermion loop interacting
with a gauge field,
\begin{eqnarray}
\label{DiracFermions}
\Gamma \left[ A \right] &\equiv& \log \int {\cal D} \psi {\cal D} \bar{\psi} \;
	\exp \left[ i \int dx \; \bar{\psi} \left(
		i\s{D} - m
	\right) \psi
\right] \\
&=& \log \left[
\det \left( i\s{D} - m \right)
\right],
\end{eqnarray}
where we have used the familiar Faddeev-Popov trick in reverse.  Now,
following the associative property of determinants and the
anti-commuting properties of $\gamma_5$ ($\det(\gamma_5) = 1$), we can
manipulate the above expression into the second order form \cite{BernDunbar},
\begin{eqnarray}
\Gamma\left[ A \right] &=& \frac{1}{2} \left( \log \left( \det \left[
		i \s{D} - m
	\right] \right)
	+
	\log \left( \det \left[
		\gamma_5 \left( i \s{D} - m \right) \gamma_5
	\right] \right) \right)
\\ &=& \frac{1}{2} \log \det \left[
	\left( i \s{D} - m \right)
	\left( - i \s{D} - m \right) \gamma_5^2
\right]
\\ &=& \frac{1}{2} \log \det
	\left( \s{D}^2 + m^2 \right)
\\ &=& \frac{1}{2} \log \det \left(
	D^2 + m^2 + \frac{g Q}{2} \sigma^{\mu \nu} F_{\mu \nu}
\right).
\end{eqnarray}
Here we use $\sigma^{\mu \nu} = \frac{i}{2} [ \gamma^\mu, \gamma^\nu
]$. For convenience, we have simply reproduced the derivation of
Strassler \cite{Strassler}.

Despite the factor of $1/2$, this generating function can be thought
of as describing a field theory for fermions. Not the original
fermions, $\psi$ of \eqn{DiracFermions}. Instead, a new set of
fermions, $\Psi$, whose effective action is,
\begin{equation}
\label{2ndFermions}
\Gamma \left[ A \right] = \frac{1}{2} \log \int {\cal D} \Psi {\cal D}
	\Psi^\dagger \; \exp \left[
	i \int dx \; \Psi^\dagger \left(
		D^2 + m^2 + \frac{g Q}{2} \sigma^{\mu \nu} F_{\mu \nu}
	\right) \Psi
\right].
\end{equation}
The Feynman rules for such a theory can be derived from this exponent.
Following an alternative derivation, we give them diagrammatically below,
see \fig{Fig2ndOrderRules}.

\Section{A hands on approach for fermion loops}

In order that we need not consider the external properties of the new
fermions, $\Psi$, we shall concentrate here on calculations of
amplitudes involving an internal fermion loop. This closed loop may in
general be embedded in a multi-loop diagram. We shall deal with {\em
external} fermions afterwards. By reorganizing the conventional
calculation for such an amplitude, we shall show how it can be
equivalently described in terms of rules of the second order
form. This approach will enable us to apply the second order rules to
cases where the determinant formula does not apply.

\begin{figure}
\begin{center}
{}~\epsfig{file=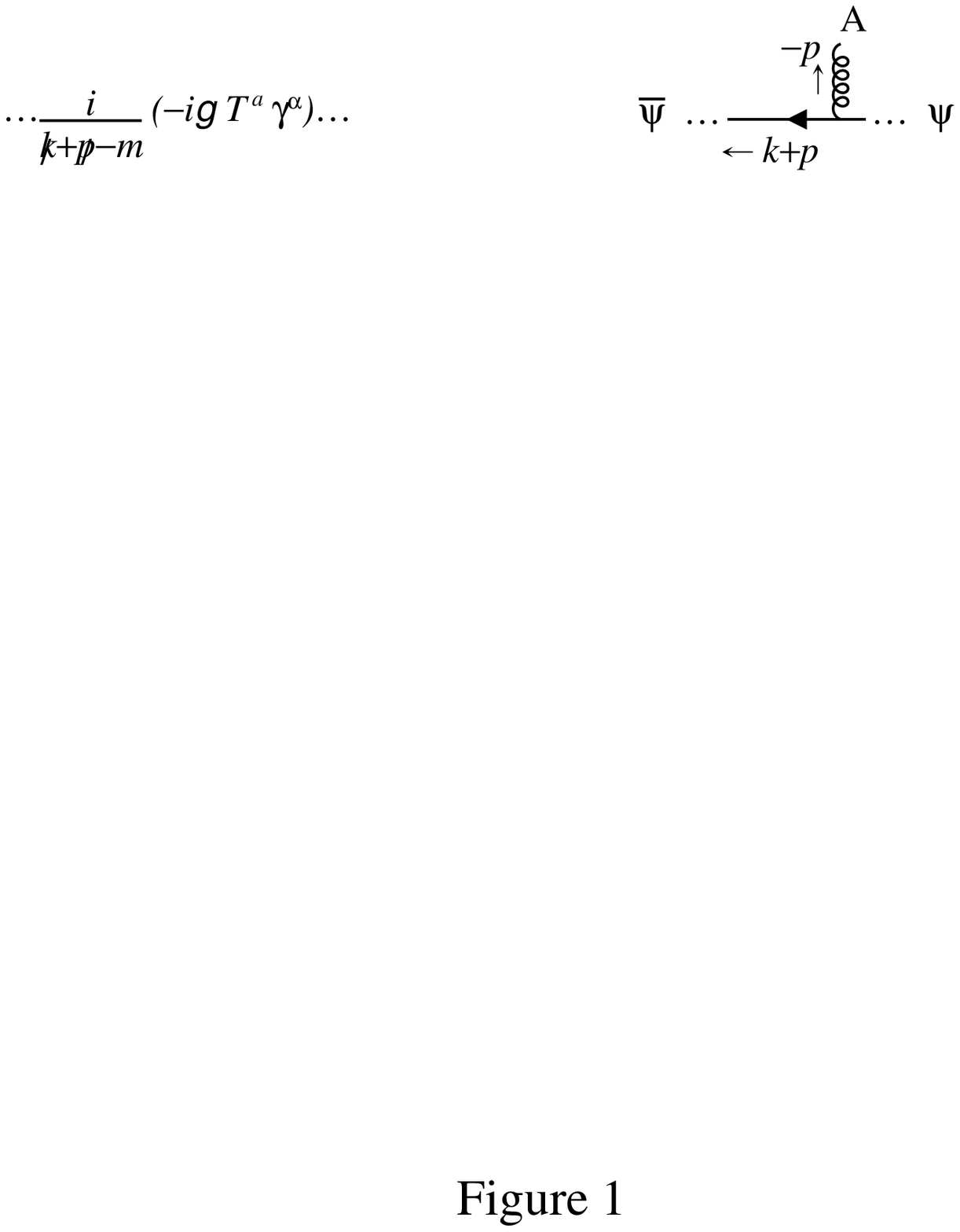,clip=}
\end{center}
\caption[]{
\label{FermionUnit}
The basic fermion unit.
}
\end{figure}
It is the case that all fermion-gauge interactions lead to the need for
evaluating the trace over some spin line. Where the spin line is
built from products of elements of the form given in \fig{FermionUnit}.
This is simply the product of a fermion propagator and a fermion-gauge
vertex.  In an abelian theory such as QED, the factor $T^a$ is the
charge of the fermion with respect to the gauge field, $Q$.  For
non-abelian theories, such as QCD, it corresponds to the appropriate
generator for the coupled gauge field. The mass of the fermion is $m$
and the gauge boson carries the momentum, $-p$, {\em away} from the vertex.

It is our intention to manipulate this unit in such a way that we
arrive at a set of equivalent Feynman rules whose principal property is
that all of the Lorentz structure is contained in the vertices.  For
brevity we write this term as $A^{\alpha,a}_{k,p}/D_{k+p}$, where,
\begin{equation}
\label{DefAD}
A^{\alpha,a}_{k,p} = i g T^a ( \s{k} + \s{p} + m ) \gamma^\alpha
\mbox{~~~~and~~~~}
D_{k+p} = i ( (k+p)^2 - m^2).
\end{equation}
By applying a Gordon-type manipulation \cite{Lam} we can write
$A^{\alpha,a}_{k,p}$ in the following form,
\begin{eqnarray}
\label{DefA}
A^{\alpha,a}_{k,p} &=& \frac{ig}{2} T^a \left(
	(2 \s{k} + 2 m + \s{p}) \gamma^\alpha + \s{p} \gamma^\alpha
\right) \\
&=& ig T^a \left(
	(k^\alpha + p^\alpha) - (-k^\alpha)
	- \frac{1}{2} [\gamma^\alpha , \s{p}]
	- \gamma^\alpha \left(
		\s{k} - m
	\right)
\right),
\end{eqnarray}
and accordingly we define,
\begin{equation}
\label{DefBandC}
B^{\alpha,a}_{k,p} = ig T^a \left(
	(k^\alpha + p^\alpha) - (-k^\alpha)
	+ i \sigma^{\alpha \beta} p_\beta
\right)
\mbox{~~~~and~~~~}
C^{\alpha,a}_k = ig T^a \gamma^\alpha \left(
	\s{k} - m
\right),
\end{equation}
such that,
\begin{equation}
\label{ASubstitute}
A^{\alpha,a}_{k,p} = B^{\alpha,a}_{k,p} - C^{\alpha,a}_k.
\end{equation}
We observe that the form of $C$ is exactly sufficient to ensure that
it generates a propagator when followed by an appropriate $A$:
\begin{equation}
\label{MakeE}
C^{\alpha,a}_{k} A_{k+q,-q}^{\beta,b} = -1 \times i (k^2 - m^2) \times -i
g^2 T^a T^b \gamma^\alpha \gamma^\beta = - D_k \times
E^{\alpha\beta,ab},
\end{equation}
where,
\begin{equation}
E^{\alpha\beta,ab} = - i g^2 \left(
\eta^{\alpha \beta}
- i \sigma^{\alpha \beta}
\right) T^a T^b.
\end{equation}
These expressions serve to define $E^{\alpha\beta,ab}$, which (after
having removed the two color matrices, $T^{a} T^{b}$) will correspond
to a color ordered four-point fermion-gauge vertex. This color
ordering arises from the fact that individual Dirac fermion loop
diagrams contribute ordered color traces to the amplitude. It has been
established elsewhere \cite{ColourOrdering}, that color ordering is a
remarkably efficient tool for use in the calculation of non-abelian
gauge theory amplitudes. To include abelian theories (QED), however,
we shall direct our attention to the sum over permuted diagrams.

We consider some general fermionic loops involving strings of these
$A/D$ objects. By re-writing selective $A$s as $(B-C)$ we will
generate an equivalent description of the fermion calculation in terms
of a new set of Feynman rules. These rules are identical to those
anticipated from the second order form of \eqn{2ndFermions}.

The generic $(n+1)$-external boson fermion loop, as derived with
conventional Feynman rules, can be written as a sum over the $n!$
orderings of boson couplings to the loop. The first of these orderings
has the following form,
\begin{equation}
\label{GenericLoop}
T_n =
- \Tr \left\{
	\frac{A_{k+ \ldots p_1,p_n}^{\alpha_n,a_n}}{D_{k+p_n \ldots + p_1}}
	\frac{A_{k+ \ldots p_1,p_{n-1}}^{\alpha_{n-1},a_{n-1}}}{
		D_{k+p_{n-1} \ldots + p_1}}
	\frac{A_{k+ \ldots p_1,p_{n-2}}^{\alpha_{n-2},a_{n-2}}}{
		D_{k+p_{n-2} \ldots + p_1}}
	\ldots
	\frac{A_{k+p_n \ldots +p_1,-p_n \ldots -p_1}^{\alpha_0,a_0}}{D_k}
\right\}
\end{equation}
where we have neglected the integration over the loop momentum. The
minus sign is the familiar fermion {\em loop} factor.
\begin{figure}
\begin{center}
{}~\epsfig{file=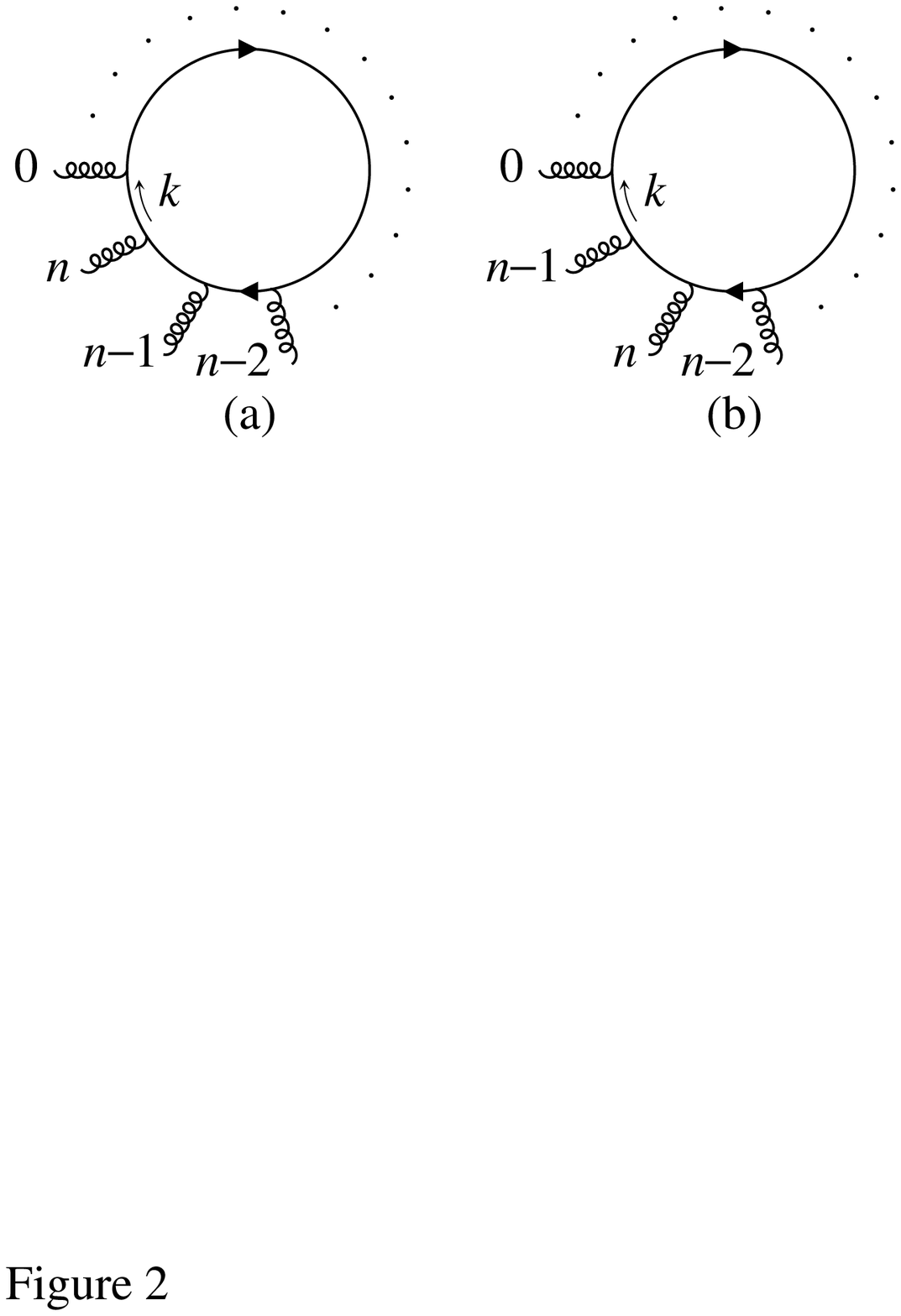,clip=}
\end{center}
\caption[]{
\label{FigNpoint}
A generic fermion loop, as described in \eqn{GenericLoop}. We label
the legs from $0$, such that the loop momentum flowing into the
$0$-vertex along the fermion line is $k$, and $p_0 = -
\sum_{i=1}^n p_i$.  }
\end{figure}
Replacing the first of the $A$s in this trace with \eqn{ASubstitute}
leads to,
\begin{eqnarray}
\lefteqn{T_n =} \nonumber \\
    &&- \Tr \left\{
	\left(
\frac{B_{k+ \ldots p_1,p_n}^{\alpha_n,a_n}}{D_{k+p_n \ldots + p_1}}
\frac{A_{k+ \ldots p_1,p_{n-1}}^{\alpha_{n-1},a_{n-1}}}{
	D_{k+p_{n-1} \ldots + p_1}}
+
\frac{E^{\alpha_n\alpha_{n-1},a_n a_{n-1}}}{D_{k+p_n \ldots + p_1}}
\right)
	\frac{A_{k+ \ldots p_1,p_{n-2}}^{\alpha_{n-2},a_{n-2}}}{
		D_{k+p_{n-2} \ldots + p_1}}
	\ldots
	\frac{A_{k+p_n \ldots +p_1,-p_n \ldots -p_1}^{\alpha_0,a_0}}{D_k}
\right\}. \nonumber \\
\end{eqnarray}
This pattern of replacement can be repeated: working our way from left
to right we substitute for the left-most $A$ with $(B-C)$ and then
apply \eqn{MakeE}. What remains once we reach the final $A$ is a
series of products of $B$s and $E$s plus a different series of $B$s
and $E$s that are truncated with a $C_{k+p_n \ldots
+p_1}^{\alpha_0,a_0}$. We note that all of the latter terms carry a
leading minus sign. For example, in the $n=1$ (self-energy) case we
obtain,
\begin{equation}
T_1 = - \Tr \left\{
\frac{B_{k,p_1}^{\alpha_1,a_1}}{D_{k+p_1}}
\frac{B_{k+p_1,-p_1}^{\alpha_0,a_0}}{D_{k}}
+ \frac{E^{\alpha_1 \alpha_0,a_1 a_0}}{D_{k+p_1}}
- \frac{B_{k,p_1}^{\alpha_1,a_1}}{D_{k+p_1}}
\frac{C_{k+p_1}^{\alpha_0,a_0}}{D_{k}}
\right\}.
\end{equation}
It is desirable to get rid of terms that contain the object $C$. And
to do this we appeal to the cyclic property of the trace. We shift the
trailing $C$ of the offending term to the left and then apply the rule
that any $B$ immediately to the right of a $C$ can be replaced by $(A
+ C)$. Here this corresponds to,
\begin{equation}
\label{TheCs}
- \Tr \left\{
-
\frac{C_{k+p_1}^{\alpha_0,a_0}}{D_{k}}
\frac{B_{k,p_1}^{\alpha_1,a_1}}{D_{k+p_1}}
\right\} = - \Tr \left\{
\frac{E^{\alpha_1 \alpha_0,a_1 a_0}}{D_{k}}
-
\frac{C_{k+p_1}^{\alpha_0,a_0}}{D_{k}}
\frac{C_{k}^{\alpha_1,a_1}}{D_{k+p_1}}
\right\},
\end{equation}
where again we have applied \eqn{MakeE}. By comparing \eqns{\ref{DefA}}
and (\ref{DefBandC}), it is simple to see that the cyclic property of
the trace guarantees that,
\begin{equation}
\Tr \left\{
C^0 C^n C^{n-1} C^{n-2} \ldots
\right\} = \Tr \left\{
A^n A^{n-1} A^{n-2} \ldots A^0
\right\},
\end{equation}
which means that the $-CC$ term, of \eqn{TheCs}, is exactly $-T_1$ and
thus we have that,
\begin{equation}
T_1 = - \frac{1}{2} \Tr \left\{
\frac{B_{k,p_1}^{\alpha_1,a_1}}{D_{k+p_1}}
\frac{B_{k+p_1,-p_1}^{\alpha_0,a_0}}{D_{k}}
+ \frac{E^{\alpha_1 \alpha_0,a_1 a_0}}{D_{k+p_1}}
+ \frac{E^{\alpha_1 \alpha_0,a_1 a_0}}{D_{k}}
\right\}.
\end{equation}
The factor of $1/2$ is quite general---for arbitrary $n$.  For this
simple example the last two terms are actually identical (this follows
from the invariance of the appropriate loop integral under the change
of loop momenta, $k \rightarrow -k-p_1$, and the cyclic property of
the trace).  Quite generally, however, we can write the sum of two
$E$s with their indices reversed as,
\begin{equation}
G^{\alpha\beta,ab} =
E^{\alpha_1 \alpha_0,a_1 a_0} + E^{\alpha_1 \alpha_0,a_1 a_0} =
- i g^2 \left(
\eta^{\alpha \beta} \left\{ T^a , T^b \right\}
- i \sigma^{\alpha \beta} \left[ T^a , T^b \right]
\right).
\end{equation}

Now, since we are attempting to create a new set of Feynman rules we
are at liberty to {\em mix} contributions from different first order
Feynman diagrams. To this end we consider the joint contribution of
\figs{\ref{FigNpoint}a and \ref{FigNpoint}b}. Amongst the many $B$ and
$E$ terms that arise from the replacement of $A$s, there are a series
of terms that are identical except for the ordering of the indices on
a single $E$ object. These terms correspond to the relative reversal
of two coupled bosons, as is depicted in the figures.

\begin{figure}
\begin{center}
{}~\epsfig{file=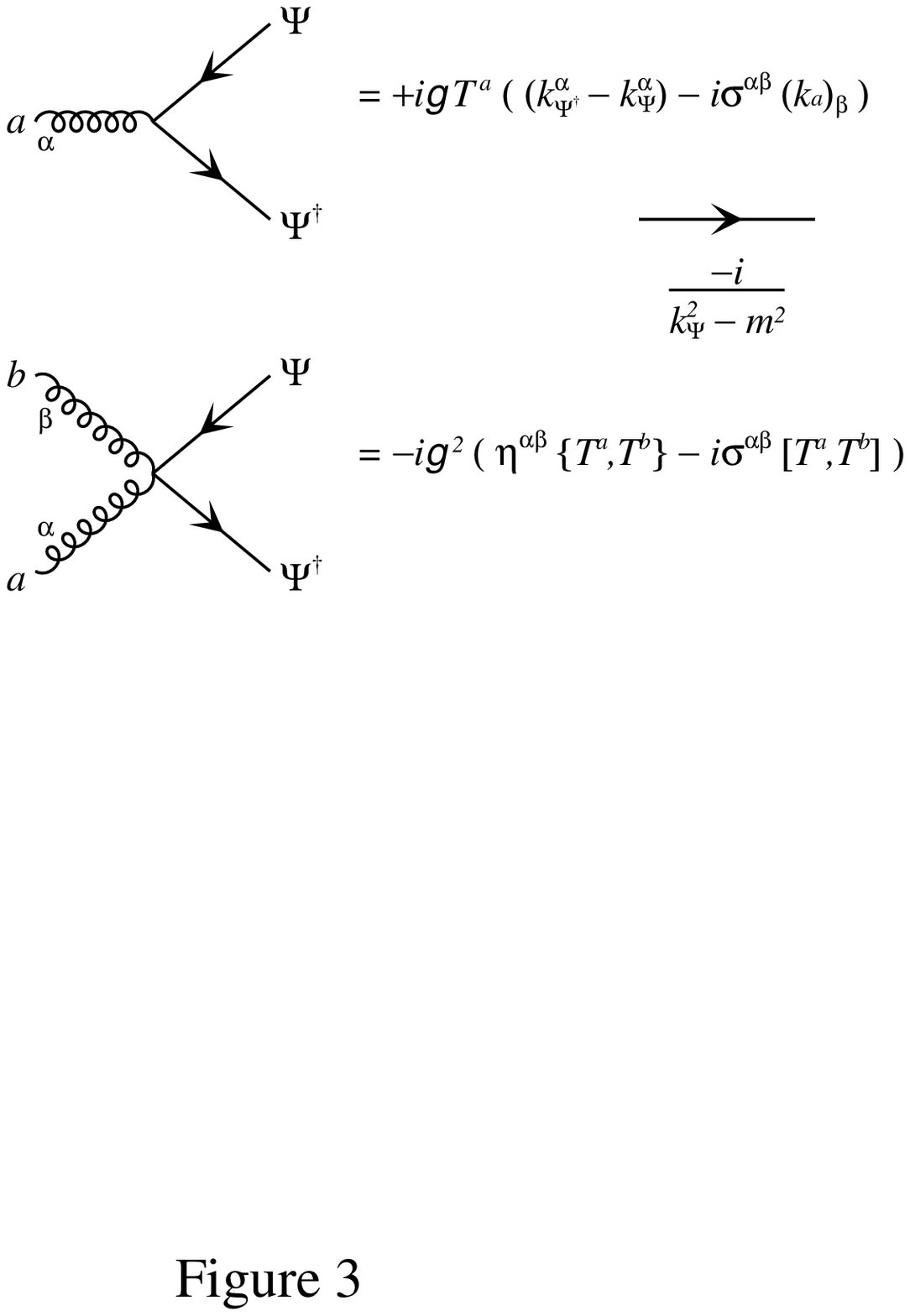,clip=}
\end{center}
\caption[]{
\label{Fig2ndOrderRules}
The second order rules for calculations of fermion-gauge interactions.
Care should be taken when using these rules to ensure that the trace
is taken over all terms and not just those containing a
$\sigma^{\alpha\beta}$. All momenta are directed outwards, which
accounts for the sign difference between the three point
$\sigma^{\alpha\beta}$ contribution and that of the $B$ in
\eqn{DefBandC}. When one traces over a closed fermion loop one should
also multiply by $1/2$. For internal closed fermion loops there is the
familiar additional factor of $-1$. See the text for the application
of these rules to processes involving external fermions.  We note that
by setting $\sigma^{\alpha\beta}\rightarrow 0$ and $i \rightarrow -i$,
these rules become those for a gauged-complex scalar field.  }
\end{figure}
It follows that any physical process described by a trace over
terms of the form $A/D$, can equally well be described by a trace over
terms built from the objects, $B$, $D$ and $G$. These {\em new} rules
for Feynman diagram calculation are given in \fig{Fig2ndOrderRules}.
We note that in the simplifying case of QED, $T^a \rightarrow Q$,
the four-point vertex reduces to,
\begin{equation}
G_{\rm QED}^{\alpha\beta} = - 2 i (gQ)^2 \eta^{\alpha \beta}.
\end{equation}
We may immediately identify these rules with those derivable from the
second order determinant, \eqn{2ndFermions}. We note that the
direction of the $\Psi$ fermion arrow is identical to that of the
first order fermion, $\psi$. That is to say, we label the head of the
fermion line by $\Psi^\dagger$. It is also worth noting that the mass
of the fermion is now only present in the denominator of the
propagator and it no longer complicates the Lorentz structure of the
Feynman rules.

\Section{A spinorial extension for external fermions}

The reorganization of Feynman rules described above is crucially
dependent on the fact that we can write the first order rules for
individual diagrams as a product of the unit-terms in
\fig{FermionUnit} ($A$s and $D$s). It is not immediately apparent that
this is the case for processes with external fermions, where the
associated amplitudes are constructed from one fewer propagators than
vertices, sandwiched between a pair of spinors. That is to say, the
leading spinor is immediately followed by a first order vertex and not
a propagator, so the string of objects sandwiched between the spinors
may not immediately be rewritten as a string of $A$s and $D$s.

At the level of the amplitude squared we should have confidence that
the process may be described with respect to the second order
rules: the loop contribution and the squared amplitude are directly
related by the optical theorem. Indeed at this level, we create a
trace by moving the trailing $u$ (or $v$) spinor of the amplitude to
precede the leading $\bar{u}$ (or $\bar{v}$) spinor of the conjugate
amplitude. Summing over the spins of the fermions we replace the
associated spinor products as follows,
\begin{equation}
\sum_{i} u_i(p,m) \bar{u}_i(p,m) = \s{p} + m
\mbox{~~~~and~~~~}
\sum_{i} v_i(-p,m) \bar{v}_i(-p,m) = - \left( \s{p} + m \right)
\end{equation}
(where we give the argument momenta of the spinors relative to the
direction of the arrow on the fermion line).  These energy projection
matrices are formed in exactly the correct positions to construct $A$
objects from the problematic first order vertices; at the head of the
amplitude and conjugate amplitude.

The absence of a propagator in the denominator, $1/D$, at the point in
the trace where the external fermions are represented ensures that no
$E$s are constructed across the cut.  Hence, this reallocation of
spinors is conveniently factorizable into two parts that we may choose
to label as a second order amplitude and conjugate
amplitude. Unfortunately, this procedure leads to a representation for
the amplitude that is a $4\times 4$ matrix, as compared to the more
efficient complex number of the conventional approach. Further we have
lost all of the spinor information for the fermions. A more useful
approach, which we shall now describe, is to effectively insert the
appropriate energy projection matrix after the leading spinor and so
retain a simple scalar representation for the amplitude.

By inserting an energy projection matrix, we can write any amplitude
involving massive first order fermions in the following way,
\begin{equation}
\bar{u}_j(k,m) \Gamma \ldots =
\frac{\bar{u}_j(k,m)}{2m} \; \left( \s{k}+m \right) \Gamma \ldots
\end{equation}
(Note, antifermion spinors, $v_j(p,m)$, have momentum arguments that
flow against the direction of the fermion arrow so in terms of the
$A$s they behave identically to the fermions but with an overall
factor of $-1$ attached.) It follows, in terms of the objects
introduced above, that each amplitude takes the form, $-\bar{u} A
(A/D) \ldots u /2m$. Equivalently, this may be written as a spinor
sandwich where the filling is constructed from the second order rules
of \fig{Fig2ndOrderRules}. The factor of $1/2$ encountered when we
dealt with closed loops is not required here because the trailing
$u_j(k,m)$ annihilates any $C_k$ that precedes it---cf. the discussion
around \eqn{MakeE}. Operationally, we generate the factor of $1/2$ by
only using the fermion or antifermion contribution to the field at the
ends of the fermion line (depending on whether it is in the initial or
final state).

Much simplification in the calculation of field theory amplitudes has
been achieved with the use of spinor-helicity methods
\cite{SpinorHelicity,KleissStirling,XuZhangChangEtc,ManganoParke}. It
is desirable to see if and how such techniques may be applied with
respect to the {\em second order formalism} for massless fermions.

The trivial rewriting above for the massive fermion case collapses in
the massless limit, $m\rightarrow 0$. Instead we may adopt a helicity
basis for the spinors (in the notation of
\cite{XuZhangChangEtc,ManganoParke}) and introduce a new null
reference momentum, $x^\mu$, which is not coincident with that of the
initial spinor in the fermion amplitude. That is to say, $x\cdot
p_{\bar{\psi}} \ne 0$. The relationship \cite{KleissStirling},
$\braket{p_{\bar{\psi}}^\pm}{x^\mp} \ket{p_{\bar{\psi}}^\pm} =
\s{p}_{\bar{\psi}} \ket{x^\mp}$, enables us to write a general
amplitude as a sum over terms of the following form,
\begin{equation}
\label{SpinorRewrite}
\bra{p_{\bar{\psi}}^\lambda} \Gamma \ket{p_{\psi}^\zeta} = \frac{
	1
}{
	\braket{x^{-\lambda}}{p_{\bar{\psi}}^\zeta}
}
	\bra{x^{-\lambda}} \s{p}_{\bar{\psi}} \Gamma \ket{p_{\psi}^\zeta} =
-\frac{1}{\braket{x^{-\lambda}}{p_{\bar{\psi}}^\zeta}}
	\bra{x^{-\lambda}} A_{p_{\bar{\psi}}} \frac{A}{D}
	\ldots \ket{p_{\psi}^\zeta},
\end{equation}
and once again we may re-express such an amplitude in terms of the
rules in \fig{Fig2ndOrderRules} sandwiched between the arbitrarily
chosen $x$-spinor and the original trailing $p_\psi$-spinor. The
choice of $x^\mu$ is as arbitrary as that made for each of the
reference momenta used to construct polarization vectors. In other
words, it is only required to be non-coincident to the original
spinor's momentum, $p_{\bar{\psi}}^\mu$. Note, as with the
polarization vectors, we cannot choose different $x$s for each of the
separate Feynman diagrams since we have mixed diagrammatic
contributions to obtain the second order rules.

An obvious choice for $x^\mu$ is that of the trailing spinor, here
given as $p_\psi^\mu$. As written out in \fig{Fig2ndOrderRules}, the
vertex rules are in general the sum of two distinct pieces: a piece
which is essentially identical to the rule for a simple scalar
interaction and a uniquely fermionic {\em Lorentz}-structure carrying
term, $\sim \sigma^{\mu\nu}$, see \fig{Fig2ndOrderRules}. The choice
$x = p_\psi$, in a single stroke, has the effect of cancelling the
{\em purely} scalar contribution to the fermionic amplitude, because
$\braket{q^\zeta}{q^\lambda}=0$. Below we shall make a different
choice to account for a supersymmetric identity at a diagrammatic
level.

\Section{An example: a supersymmetric identity}

Since the second order rules bear such a close resemblance to the
Feynman rules for a scalar interacting with a gauge field, it is worth
investigating whether they shed light on the the supersymmetric Ward
identities \cite{SusyWardIdentities,ManganoParke}. We shall, for the
sake of example, consider the simple relationship between the
${\cal A}(-+;+\ldots-\ldots+)$ amplitudes of the $\phi^\dagger \phi\,+\,
(n-2) \gamma$ and $\bar{\psi} \psi\, +\,(n-2) \gamma$ processes at
tree level. Here the $\gamma$s represent abelian photons, $\phi$ a
complex scalar and $\psi$ a fermion. We use the convention that the
positive (negative) helicity label for the scalar refers to its
particle (anti-particle) state. The non-abelian amplitudes obey the
same identity \cite{ManganoParke}, but for the sake of simplicity we
shall only present the abelian case.

We shall need the following identity,
\begin{equation}
\label{UsefulIdent}
\left[ \s{\epsilon}^{\lambda} (p;r), \s{p} \right] =
\frac{4 \lambda}{\sqrt{2}}
\ket{p^{-\lambda}} \bra{p^{\lambda}}
{}.
\end{equation}
Here we use the polarization vector \cite{XuZhangChangEtc,ManganoParke},
\begin{equation}
\epsilon^\lambda_\mu (p;r) = \lambda \frac{
	\bra{p^\lambda} \gamma_\mu \ket{r^\lambda}
}{
	\sqrt{2} \braket{r^{-\lambda}}{p^\lambda}
},
\end{equation}
where $p$ is the physical momentum of the photon carrying helicity
$\lambda$, and $r$ is the reference momentum.

\begin{figure}
\begin{center}
{}~\epsfig{file=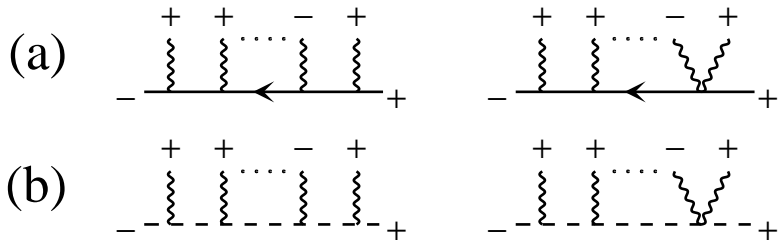,clip=}
\end{center}
\caption[]{
\label{FigSusy}
This figure represents contributions to the MHV amplitudes which enjoy
a close supersymmetric similarity. The text describes the simple
relationship between the (a) second order fermionic and (b) scalar
amplitudes.  }
\end{figure}
Now, with reference to \fig{FigSusy}, we shall compare the product of
terms used in each of the diagrams that contribute to the fermion
(\fig{FigSusy}a) and the scalar (\fig{FigSusy}b) amplitudes.  Such
amplitudes are termed maximally helicity violating (MHV), because they
are those amplitudes with the most external legs of the same helicity
that are not identically zero at tree level.

In terms of the second order rules, we can see quite quickly that the
${\cal A}^\psi(++;\ldots +)$ and ${\cal A}^\psi(-+;\ldots +)$
amplitudes vanish. The former because each of the rules is helicity
conserving and $\braket{n^+}{p_\psi^+}=0$ (this result is also obvious
when using the first order rules). The latter follows from the
following fact,
\begin{equation}
\label{FermionGone}
\bra{n^-} [\s{\epsilon}^+(p;r),\s{p}] = 0
\end{equation}
(see \eqn{UsefulIdent})---that is to say, where all the photon
helicities are opposite to that of the leading spinor the fermion and
scalar vertices are effectively identical. Choosing the arbitrary
spinor momentum at the head of the fermion line to be equal to the
momentum of the trailing spinor, $x = p_{\psi}$, gives an amplitude
proportional to
$\braket{p_\psi^-}{p_\psi^+}/\braket{p_\psi^+}{p_{\bar{\psi}}^-} =
0$. In this latter case the proportionality factor is the scalar
amplitude, ${\cal A}^\phi(-+;\ldots +)$. This can be seen to be zero
by choosing the reference momenta of every polarization vector to be
equal to momentum of the trailing spinor, $p_\psi$.  So in fact both
of these fermion amplitudes are zero independent of our choice for
$x$. This can also be seen as a consequence of supersymmetric Ward
identities \cite{SusyWardIdentities}.

For the MHV amplitudes the relationship between the scalar and
fermionic amplitudes is almost as simple as that of the ${\cal
A}(-+;\ldots +)$ case. This is because each of the three-point
vertices that couple the positive helicity photons to the fermion are
effectively just scalar vertices (see \eqn{FermionGone}). The
difference between the fermion and scalar amplitudes is due solely to
the {\em uniquely fermionic} component of the negative helicity
three-point vertex. This contribution is of the form (see the first of
the graphs \fig{FigSusy}a and \eqn{UsefulIdent}),
\begin{equation}
\frac{
	\sqrt{2} i g Q
}{
	\braket{x^-}{p_{\bar{\psi}}^+}
}
\bra{x^-} \; S_{\rm before} \; \ket{p_{(-)}^+} \bra{p_{(-)}^-} \;
S_{\rm after} \; \ket{p_\psi^+}
\end{equation}
where $p_{(-)}$ is the momentum of the negative helicity
photon. $S_{\rm before(after)}$ is the product of the
preceding(trailing) scalar-like vertices and propagators, which freely
commute with $\bra{x^-}$. Clearly the choice, $x = p_{(-)}$, removes
this contribution from the amplitude.

Using \eqn{SpinorRewrite} with the choice $n=p_{(-)}$ for the spinor
reference momentum, it becomes trivial with the {\em second order}
rules to demonstrate in a perturbative sense that each of the {\em
diagrams} contributing to these tree level MHV amplitudes (and
accordingly the amplitudes themselves) are related by,
\begin{equation}
{\cal A}^\psi(-+;\ldots - \ldots+) =
\frac{
	\braket{p_{(-)}^-}{p_\psi^+}
}{
	\braket{p_{(-)}^-}{p_{\bar\psi}^+}
}
{\cal A}^\phi(-+;\ldots - \ldots+).
\end{equation}
It is amusing to note that this diagram-by-diagram equivalence is even
independent of the choice of reference momenta for the photon
polarization vectors. At the level of the amplitude this is of course
in complete agreement with the appropriate supersymmetric Ward
identity \cite{SusyWardIdentities}.

\Section{Summary}

In this letter, we have discussed the {\em second order} formalism for
fermions and have succeeded in rewriting the Feynman rules for
gauged-fermions in a form that exposes their close similarity to
gauged-scalars. We have shown how these rules may be applied to the
calculation of both closed fermion loops and for processes involving
external fermions.  Because of the close relationship between the
second order rules and the first quantized form
\cite{Polyakov,Strassler}, this work may provide some
reference for those wishing to construct string-like rules for
external fermions.

The supersymmetric relationship between fermion and scalar {\em loops}
has been investigated previously
\cite{BernKosower,BernDunbar,BernEtAl} where the second order
formalism, as it applied to closed fermion loops, was employed. Here,
we have demonstrated that this similarity exists for external fermions
too. We have used it in a simple example, to make transparent the
supersymmetric Ward identity relating two MHV amplitudes, which is
something not normally obvious at the level of individual Feynman
diagrams.

Although the simple relationship presented in the above example is
more complicated for other helicity configurations, it will always be
the case that a fermion amplitude has a component which is directly
proportional to a scalar one---a feature which is explicit with the
second order Feynman rules. This component is physical, insofar as it
is gauge invariant. So by choosing a different set of reference
momenta for the external gauge bosons, we may optimize the calculation
of the {\em fermionic} remainder. That is to say, such a remainder may
be calculated more efficiently on its own than when mixed up with the
rest of the amplitude. This method of breaking-up amplitudes has
already been used successfully for closed fermion loops
\cite{BernEtAl}, and can be thought of as a generalization of the
established {\em supersymmetric} techniques
\cite{Kunszt,ManganoParke}.


\section*{Acknowledgments}
It is a pleasure to thank Zvi Bern and Lance Dixon for reading the
manuscript and their many useful suggestions. Further I thank the
following people for much helpful discussion, Adrian Askew, Daniel
Cangemi, Nigel Glover, David Kosower and Duncan Morris. This work was
supported by the DOE under grant DE-4-444025-PE-22409-2.

\end{document}